# Turing Test and the Practice of Law:
# The Role of Autonomous Levels of AI Legal Reasoning

**Dr. Lance B. Eliot**
Chief AI Scientist, Techbruim; Fellow, CodeX: Stanford Center for Legal Informatics
Stanford, California, USA

**Abstract**

Artificial Intelligence (AI) is increasingly being applied to law and a myriad of legal tasks amid attempts to bolster AI Legal Reasoning (AILR) autonomous capabilities. A major question that has generally been unaddressed involves how we will know when AILR has achieved autonomous capacities. The field of AI has grappled with similar quandaries over how to assess the attainment of Artificial General Intelligence (AGI), a persistently discussed issue among scholars since the inception of AI, with the Turing Test communally being considered as the bellwether for ascertaining such matters. This paper proposes a variant of the Turing Test that is customized for specific use in the AILR realm, including depicting how this famous "gold standard" of AI fulfillment can be robustly applied across the autonomous levels of AI Legal Reasoning.

## 1   Background and Context

Artificial Intelligence (AI) is increasingly being applied to law and legal tasks amid attempts to bolster AI Legal Reasoning (AILR) autonomous capabilities [1] [5] [11] [17]. The use of Machine Learning (ML) and Deep Learning (DL) has significantly aided in making improvements and advances in AILR systems [27] [31]. Also, ML/DL in Natural Language Processing (NLP) has made tremendous strides in computational fluency and semantic analysis performance that has bolstered the use of LegalTech for e-Discovery, contract creation, searches of a large corpus of court cases, and the like [14] [26] [40].

A major question that has generally been unaddressed involves how we will know when AILR has achieved autonomous capacities. So far, AI as applied to the legal profession has primarily consisted of aiding or supporting the legal work of human lawyers but has not reached the capability of being able to autonomously perform legal tasks. A base assumption is that inexorably there will be advances made in AI that will boost AILR systems and ultimately transcend them into having autonomous capacities, but there does not yet exist any bona fide and nor rigorous means to viably attest to whether such AILR autonomy has been achieved [44].

Without an acknowledged and universally accepted method or means of attesting to AILR autonomy, a vacuum remains that will likely stoke false claims and confound those within the law industry and those outside the legal field. Vendors providing AILR systems will continue to be able to assert they have been able to develop autonomous AILR, doing so with shallow assertions buoyed-up by whatever obtuse measures they wish to stake such a claim on. Likewise, AI and law researchers that are striving to make scholarly foundational advances in AILR will lack any viable means to discern the pace and scope of progress in creating AILR autonomous functionality.

The overarching field of AI has grappled with alike quandaries concerning how to assess the attainment of Artificial General Intelligence (AGI). AGI refers to the goal of seeking to achieve AI that can be on par with human intelligence and thus convincingly demonstrate the same caliber and depth of reasoning as that of human cognition. Discussion and debates over how to assess whether AGI has been attained have permeated the field of AI since its very inception. Generally, the Turing Test [52] has commonly been considered the bellwether for ascertaining such matters and is known worldwide as a method or approach to the testing of AI, having been devised by the famous mathematician Alan Turing in 1950 [39] [41] [47] [51].

This paper proposes a variant of the Turing Test that is customized for specific use in the AILR realm, including depicting how this renowned "gold standard" of AI fulfillment can be robustly applied to AILR. Also, the paper makes use of a framework of autonomous levels of AI Legal Reasoning [20] [21] [24], indicating how the Turing Test applies at each successive level of AILR autonomy. The proposed grid and discussion are intended to contribute to the study of AI & Law as this burgeoning realm seeks to identify and mature a method or means to suitably determine and formally assess AI Legal Reasoning autonomous systems.

This paper consists of five sections:
- Section 1: Background and Context
- Section 2: Key Elements of the Turing Test
- Section 3: Autonomous Levels of AI Legal Reasoning
- Section 4: Turing Test Grid Integrating Autonomous Levels of AILR
- Section 5: Additional Considerations and Future Research

In Section 1, an overall background about the Turing Test is provided. Section 2 then goes into a further in-depth analysis of the Turing Test and identifies the key elements involved. In Section 3, an overview is provided on the autonomous levels of AI Legal Reasoning, which is crucial to then understanding Section 4. Section 4 proposes a grid that aligns the Turing Test elements with the autonomous levels of AI Legal Reasoning. Section 5 is a discussion of additional considerations and also offers suggested avenues for future research on these matters.

**1.1 Overview of the Turing Test**

Noted mathematician Alan Turing proposed the Turing Test in 1950 when trying to address the question of whether machines can think [52]. He was desirous of avoiding getting mired in debates about what thinking consists of, which can readily get hindered in the murky and unknown underpinnings of the brain and cognition. Note that even today, some 70 years later, the means of how we think are still largely undetermined.

The overarching notion by Turing was to treat thinking as a black box and thus not need to ascertain the internal mechanisms. He conceived of a testing approach that would avert relying upon how thinking is derived, and instead be aimed at the resultant behavior that thinking produces. He also wanted to separate the physical aspects of thinking from the intellectual aspects. In essence, a thinking machine does not necessarily need to have a human body or any semblance of a body, and might instead be encapsulated in a computer-based system that does not showcase itself in a human-like way (i.e., it does not need to be a robot that has the appearance of a human figure).

Some immediately criticized the Turing Test for averting the ongoing question of mind-body, whereby some theorists suggest that the human mind and the act of thinking are intertwined, and it is not possible to separate the two [41] [55]. This criticism though is addressed by the simple fact that the test as devised would presumably lead to a failure on the part of the AI if indeed a mind-body composition is an absolute requirement for the act of thinking since the AI would assuredly be unable to demonstrate thinking as it has no such body or encasement included.

Per Turing [52], he stated that "It is natural that we should wish to permit every kind of engineering technique to be used in our machines," and thus he wanted to devise a test that would not limit how a thinking machine could be developed. He also anticipated the retort that if any machine is allowed, potentially a person might be cloned via biological techniques, and this "machine" considered a form of AI due to it being "engineered" into existence. Turing [52] remarked that "To do so would be a feat of biological technique deserving of the very highest praise, but we would not be inclined to regard it as a case of 'constructing a thinking machine.'"

Therefore, it is assumed that for the sake of the Turing Test, a reasonableness perspective be taken about the AI and how it is embodied and that it somehow is considered to be commonly denoted as a "machine" and not a biological person (trying to definitively define the meaning of "machine" can in itself be a significant burden).

The Turing Test that he devised consists of a person that takes the role of conducting an interrogation, asking questions of two subjects or participants, one being a human and the other being a (potential) thinking machine, and neither is visible to the interrogator. Imagine that the two subjects are hidden behind a curtain on a stage and that the interrogator can only interact indirectly via speaking or writing a message to them but cannot see them directly. This hiding of the subjects aids in what otherwise would be a rather perfunctory exercise of merely looking at the participants and visually ascertaining which is the human and which is the machine (assuming that the machine is not a robot fashioned to look identically like a human).

The interrogator does not know beforehand which of the two is the human and nor which of the two is the AI. For sake of convenience, label one of them as X and the other as Y. The interrogator asks questions or makes queries of the X and Y, and at some point, ascertains that the effort should be concluded. Upon so ending the effort, the interrogator is then to state whether X is the human or whether Y is the human, which alternatively could be stated by indicating whether X is the AI or whether Y is the AI.

Turing referred to this test as the "imitation game" since it involves the AI attempting to imitate the human, though this might or might not be the intention per se of the AI. It could be that the AI has been devised to be a thinking machine, and thus it "mimics" the act of human thinking. Whether this kind of thinking is the same as human thinking is a longstanding open debate, therefore that can be somewhat sidestepped by suggesting that the AI is mimicking human thinking, regardless of whether it is, in fact, identical in how it thinks or does so in an entirely different manner.

Many prefer to refer to the imitation game as the Turing Test, rather than mentioning that it is a game, which perhaps undermines the cogent value of the approach. When considering games or contests, we might be quick to dismiss them as nothing of worthwhile consequence. Some suggest too that the Turing Test is more so an experimental arrangement, and thusly refer to the human participant and the AI as subjects, akin to the phrasing used in scientific experiments.

The aim of the Turing Test is that if the interrogator is unable to differentiate between the two subjects, presumably the AI is thusly indistinguishable from the human, in terms of thinking, and thus we can conclude that the AI has achieved the equivalence of human intelligence. This greatly simplifies the seemingly intractable problem of trying to define what human intelligence consists of. If the AI can demonstrate intelligence to the same degree as a human, it can be said to be a thinking machine.

When pursuing the consideration of human intelligence, Turing envisioned that a question and answer dialogue would be a key means for the integrator to try and separate the chaff from the wheat, so to speak, and assumed that the interrogator is sufficiently capable undertaking the interrogation, and ultimately able to reach a reasoned conclusion about which is the human and which is the AI. Turing suggested that the interrogator might ask the subjects to write poetry, or play chess, or do whatever kinds of mentally engaging tasks that might be deemed worthwhile for purposes of conducting the test.

The Turing Test has been pervasive in the field of AI over the many years since it was first proposed and continues today as a commonly referred to "standard" of how to assess the achievement of machine-based human-like intelligence [3] [4] [39]. Various tournaments have taken place using the Turing Test, along with prizes offered for being able to devise AI that can succeed at winning a Turing Test. It is crucial though to realize that none of these various Turing Tests were of the variety envisioned by Turing, and instead are extremely reduced versions, oftentimes limiting the test to a pre-determined scope or a set time limit. Thus, there is not yet any AI that has been able to successfully win or pass a Turing Test of an unencumbered nature that was robustly attempting to ascertain general intelligence.

Some have used the Turing Test to examine AI in specific disciplines. For example, in the medical field, there are AI systems that purportedly exhibit human intelligence capacities when analyzing an X-ray or MRI, and thus a type of Turing Test can be set up to try and determine the veracity of such medically specialized claims [42] [57]. Note that this is not the same as determining general intelligence and instead of a focus on so-called narrow AI.

In the field of law, efforts to apply AI to legal reasoning have also at times referred to the Turing Test, proposing that it be utilized for ascertaining the capabilities of AI LegalTech systems. Such suggestions have not been laid out in specified detail and are typically an overall reference to the importance and potential applicability of using the Turing Test in the application of AI to the law [41] [45].

In an unusual and intriguing perspective about the Turing Test, Reinbold [44] discusses the Turing Test in the context of patents. Currently, the United States does not allow AI to be considered a patent inventor, but some argue that AI ought to be permitted to hold a patent. Reinbold suggests that the Turing Test could be used to aid in deciding whether AI should be eligible for being granted a patent [44]: "Principally, AI that passes the Turing Test constitutes 'inventive AI' and likely produces unpatentable inventions under 35 U.S.C. § 103. In contrast, AI that fails the Turing Test permits user control and influence over the inventive process and may result in patentable 'AI-assisted inventions.'" In short, if the AI passes the Turing Test, it cannot be granted a patent under existing provisions, while if it fails then it could potentially be considered an AI-assisted invention.

Overall, there is a gap or opening within the field of AI and law that leaves unstated how we will know when AI has reached a sufficiency of being able to practice law and thus might be permitted to do so, autonomously rather than via working solely on a human attorney-assisted basis. This paper proposes that the Turing Test be tailored to the discipline of law, and by doing so would provide a means to assess AI applications purporting to perform legal reasoning.

In the next section, this paper identifies the key elements involved in the Turing Test and discusses how those elements can be tailored or customized to the assessment of AI-based Legal Reasoning.

**2 Key Elements of the Turing Test**

In this section, the key elements of the Turing Test are identified. An explanation for each key element is provided. This will be crucial for then applying these elements to the autonomous levels of AILR.

**2.1 Turing Test and Key Elements**

The key elements are depicted in the below short-form descriptors that are considered suitable for use in a grid and consist of keywords to represent each element. The key elements consist of:
- The Inquirer
- Human Participant
- AI-Based Legal Reasoner
- Queries of the Turing Test
- Answers to the Turing Test
- Rules of the Turing Test
- Potential Observers
- Conclusion Reached
- Reverse Turing Test

In the subsections, each key element will be briefly explained and explored.

**2.2 Details Underlying Key Elements**

For each of the key elements, it is foundational to explain the nature and scope of the element, doing so to ensure that each is representative of its focused intent.

**2.2.1 Element "The Inquirer"**

The person that asks the questions of the Turing Test participants is customarily known as the interrogator, which was the wording originally used by Turing in describing the overall arrangement. Since the word "interrogator" can have varied connotations associated with it, which invokes for some a semblance of antagonism or other definitional baggage, the person in the role of conducting the inquiry has oft been coined as the inquirer. There are additional wording variations such as being referred to as the evaluator, sometimes also referred to as the judge, and so on.

For purposes of this study, the word "inquirer" is utilized.

Doing so is for purposes of seeking to avoid any otherwise distracting confusion or confounding considerations about the role. The word "inquirer" is presumed to be less likely to trigger any adverse reactions about the nature of the role and thus is considered a relatively neutral phrasing. Regardless of the phrasing chosen, the role is still the same role as originally envisioned.

One question to be considered about the inquirer is whether this is denoted as one person or whether it could be more than one. The original portrayal implied it would be one person, though this was not an aspect that garnered particular attention or was raised as a potential consideration in the initial arrangement.

The viewpoint taken here is that it would be feasible to have a Turing Test with more than one inquirer, which is a reasonable stance given that the role of the inquiry overall is to try and assess the full range of human intelligence and whether the AI can exhibit that entire range. It would seem problematic to assume that one person alone in the inquirer role could cover the varied breadth and depth of human intelligence, and as such, there might be multiple inquirers employed for the role. Ideally, the number of inquirers would be kept to a reasonable number and there would be cogent coordination among the inquirers too. This concept of multiple persons in the inquirer role is deserving of additional research and will be mentioned further in Section 5. When using the word "inquirer," henceforth herein this is intended to indicate the role of the inquirer and for which it might be one or more persons.

Another aspect of the inquirer role is that it is a multi-faceted role. As originally described, the inquirer asks questions of the participants, acting in a somewhat prosecutorial manner, and simultaneously is gauging the answers, acting in a somewhat judicial manner, along with ultimately rendering a decision as to the outcome of the Turing Test. Whether this is an unduly overloaded role has been previously questioned. Likewise, this brings up the corresponding concern that having one person that serves essentially as a mix of a prosecutor, judge, and jury would seem inherent to have the undue potential for problematic sway including incurring cognitive biases as the inquirer (an inquirer might be swayed by their own choice of questions, whereas if there was a separate evaluator they might independently be better served at assessing the answers of the participants, and so on). This matter is not addressed per se in this study and merely noted as a consideration about the nature of the Turing Test and for purposes of potentially spurring further research on the matter.

All told, the person that undertakes the inquirer role is notably significant since how the person conducts the Turing Test is tantamount to shaping the worth of the effort and its outcome. Someone that is insufficiently capable in this role would undeniably undercut the significance of the Turing Test.

In the overarching Turing Test, the inquirer is covering all facets of general intelligence. For purposes of the Turing Test utilization in the context of this study, the inquirer is focused on the discipline of legal reasoning.

**2.2.2 Element "Human Participant"**

The human participant is the barometer against which the AI system is being compared, and therefore it is essential to the Turing Test that the human participant be sufficiently capable in this role.

As similarly discussed in the prior subsection about the inquirer, the human participant was originally depicted as one person rather than being multiple people at once. The underlying question arises regarding whether it is reasonable to expect that one person alone would be capable of serving in this crucial barometer capacity. As such, it is conceivable that the human participant could consist of one or more humans and that they would need to be coordinated in their efforts thereto. This is a concept deserving of additional research and will be so mentioned in Section 5.

Another facet of the human participant is the base assumption that the human participant will genuinely perform when undertaking the Turing Test. If the human participant is insincere in their effort, it would undoubtedly undermine the nature of the testing activity. There is a counterargument sometimes made that this could also be a ploy by the AI, attempting to portray itself in a human-like manner. In that same logical vein, the human participant could attempt to masquerade as the AI, if one assumes that the AI can be so mimicked.

Yet another aspect involves whether the human participant can be equipped with the use of a computer. Purists would tend to argue that the human should be entirely unaided and be acting solely based on their own intellect. Where this comes to play would involve the aspect of the inquirer asking each of the subjects to calculate a large value, and when one of them is unable to do so or takes a long time to do so, the human participant is revealed. To solve this, the belief is that the human participant should be permitted to use a computer. But this introduces additional complications, such as if the computer is running the same AI as the AI being used for the comparator, does the Turing Test make any reasonable sense when the human participant is armed with the same AI. For purposes of this study, the viewpoint is taken that the human participant would likely need to have available some computer-based capacities due to the nature of the context, yet would need to be limited in having access to the AI per se (this is a matter mentioned further in Section 5 for future research exploration).

In the overarching Turing Test, the human participant is expected to cover all facets embodying general intelligence. For purposes of the Turing Test utilization in the context of this study, the human participant is focused on the discipline of legal reasoning.

### 2.2.3 Element "AI-Based Legal Reasoner"

The computer-based AI is the comparator to the human participant.

In the overarching Turing Test, the AI is intended to cover all facets of general intelligence. For purposes of the Turing Test utilization in the context of this study, the AI is focused on the discipline of legal reasoning and will be denoted as the AI-based Legal Reasoner.

Referring to AI overall has an implied monolithic insinuation, which should not be necessarily taken or interpreted in that manner. It could be that the AI is a federated system with numerous components that work in conjunction with each other. Note that however the AI is formed, including the underpinnings of technology used, does not bear on the Turing Test in any substantive way. The Turing Test is essentially technology agnostic and there is no indication and nor assertion as to what or how the AI is composed and undertaken.

Another aspect of the AI involves whether the AI might be devised to attempt trickery at mimicking the human participant or the nature of human responses. Some have labeled this ploy as a form of Artificial Stupidity, arising from the notion that if the AI is asked to calculate a complex equation, and arrives at an answer with fifty digits, this perhaps gives away the AI, and thus the AI might purposely act as though it only knows a few of the digits, or perhaps even offers the digits erroneously as though having made an error that a human might make. Some argue that this is entirely at the choice of the AI to decide whether to attempt and that doing so could either aid in the AI appearing to be human-like or might backfire on the AI by revealing that it is the AI and exploiting such a ploy by appearing to be dimwitted or human-like error-prone.

### 2.2.4 Element "Queries of the Turing Test"

The original establishment of the Turing Test did not specify the nature of the queries that the inquirer is supposed to ask of the human participant and the AI. Presumably, the inquirer should use their intellect to devise a sufficient series or set of questions that can achieve the end-goal of being able to ascertain whether the AI can be distinguished from the human participant. Furthermore, it might be reasonably assumed that the inquirer could devise new questions in real-time as needed, doing so in response to the answers of the AI or the human participant. This kind of interactive dialogue would seem the more likely means to try and discern the intellectual prowess of the subjects.

Some have outlined the kinds of queries that might be used in a general intelligence Turing Testing. Nonetheless, there is no universally accepted set or specification of what the queries need to be.

Per word choice, herein the word "query" or "queries" is used, rather than words such as "questions" or the "inquiries," though those other phrasings are equally applicable and considered interchangeable for purposes herein.

In the overarching Turing Test, the queries are expected to cover all facets embodying general intelligence. For purposes of the Turing Test utilization in the context of this study, the queries are focused on the discipline of legal reasoning.

### 2.2.5 Element "Answers to the Turing Test"

The answers that are to be provided by the human participant and by the AI are presumed to be completely open-ended, meaning that their respective answers are whatever answers they wish to provide. It is then up the inquirer to decide whether the answers are appropriate and whether the answers are sensible or nonsensical, etc.

In the overarching Turing Test, the answers are expected to cover all facets embodying general intelligence. For purposes of the Turing Test utilization in the context of this study, the answers are presumed to be focused on the discipline of legal reasoning.

A longstanding question about legal reasoning is the degree to which law and legal reasoning involve and depend upon general intelligence, such that there might be little means of separating legal reasoning from general intelligence. In that sense, it could be asserted that the Turing Test in a legal context has no choice but to also involve the use of general intelligence, and therefore it is somewhat misleading to suggest that a Turing Test for legal reasoning is solely and exclusively only about the law and legal reasoning. This significant point is worthwhile to keep in mind.

**2.2.6 Element "Rules of the Turing Test"**

There are no established rules for the Turing Test, other than the general semblance of the inquirer opting to ask queries of the human participant and the AI, doing so in whatever manner the inquirer deems to do so. In other words, the inquirer does not need to alternate between the subjects, does not need to be balanced in asking questions, and so on. This is left entirely up to the discretion of the inquirer.

In theory, the inquirer could ask queries of only one of the subjects and opt to not ask any of the other. Furthermore, the inquirer could ask just one question and offer no other questions for the subjects. Since it would be a seeming undermining of the Turing Test for the inquirer to take such a stance, it has been proposed that there should be some explicitly stated rules associated with the Turing Test.

In the case of attempts at undertaking the Turing Test, there have been various rules sketched, though they have tended to be narrow and overly specific. For example, suppose a Turing Test is undertaken that stipulates the entire testing period will be five minutes in length. This does not seem a sufficiently long enough period to allow for a properly undertaken inquiry, and thus the resulting outcome would be specious or certainly suspect.

In the overarching Turing Test, rules would presumably be crafted aiming to cover all facets embodying general intelligence. For purposes of the Turing Test utilization in the context of this study, rules are presumed to be focused on covering the discipline of legal reasoning.

**2.2.7 Element "Potential Observers"**

In the original description of the Turing Test, there is no delineation of whether there might be observers involved in the Turing Test. Essentially, it is not a topic particularly brought up or considered. Subsequently, it has been envisioned that there would seem to be value in having observers, without which otherwise the nature of the Turing Test might be perceived as less viably undertaken and ultimately discounted.

Some assert that the inclusion of observers might impact the Turing Test and alter the results, somehow skewing the effort. Others point out that the observers could be kept astray of the matter and nonetheless still be able to observe the effort. If done properly, it can be argued that the inclusion of observers has no material effect on the Turing Test itself, while at the same time perhaps achieving an acceptance or acknowledgment of the result due to the allowance for having observers.

In the overarching Turing Test, observers would be principally anyone having an interest in general intelligence. For purposes of the Turing Test utilization in the context of this study, observers are presumed to be focused on having a particular interest in the discipline of legal reasoning.

**2.2.8 Element "Conclusion Reached"**

The primary outcome of the Turing Test consists of the inquirer declaring which of the subjects is the AI. If the inquirer correctly states which is the AI, presumably the AI has failed at being able to showcase the equivalence of human intelligence and somehow given itself away, thus, "failing" the Turing Test. If the inquirer is unable to state which is the AI, presumably the AI has been able to showcase the equivalence of human intelligence and thus "succeeded" in passing the Turing Test.

There are numerous qualms about this simplistic standpoint. Suppose for example that the inquirer merely flips a coin to ascertain which of the subjects is the AI. In that case, would the AI have "succeeded" if the coin toss failed to select the AI, and would the AI have "failed" if the coin toss perchance selected the AI? This certainly does not seem suitable. Another concern is that doing the Turing Test perhaps once, and then declaring a failure or success does not seem especially valid, and perhaps it ought to be done repeatedly until some level of repeated efforts provides a more substantive basis for rendering a result.

In the overarching Turing Test, the conclusion reached would be whether the AI has apparently demonstrated general intelligence. For purposes of the Turing Test utilization in the context of this study, the conclusion reached is whether the AI has achieved sufficiency in the discipline of legal reasoning.

### 2.2.9 Element "Reverse Turing Test"

The traditional or conventional Turing Test has been described in these subsections. A variant known as the Reverse Turing Test has been identified in the literature and variously defined. One variant is that the Reverse Turing Test consists of the inquirer having to identify which of the subjects is the human, rather than which of the subjects is the AI. This of course does not appear to be demonstratively different than the conventional approach since by the act of identifying which is the AI, by default the assumption is that the other subject is indeed the human participant. Nonetheless, some assert that the focus on trying to identify the human participant rather than the AI is a notable difference and therefore merits its special attention as an approach to the Turing Test.

Another meaning for a Reverse Turing Test consists of the human participant attempting to masquerade as the AI. The basis for doing so is sometimes attributed to a software development technique called the Wizard of Oz, whereby a software developer pretends to be the computer and responds to human end-users, seeking to ferret out what kinds of interaction the human end-users are desirous of having, and then programming the computer system accordingly.

For purposes of the Turing Test utilization in the context of this study, the Reverse Turing Test is included as a form of completeness of coverage, without stipulating or assessing the value of the approach.

## 3.0 Autonomous Levels of AI Legal Reasoning

In this section, a framework for the autonomous levels of AI Legal Reasoning is summarized and is based on the research described in detail in Eliot [20].

These autonomous levels will be portrayed in a grid that aligns with the Turing Test key elements identified in the prior section of this paper, and thus it is useful to first explain what each of the autonomous levels consists of.

The autonomous levels of AI Legal Reasoning are as follows:

Level 0: No Automation for AI Legal Reasoning
Level 1: Simple Assistance Automation for AI Legal Reasoning
Level 2: Advanced Assistance Automation for AI Legal Reasoning
Level 3: Semi-Autonomous Automation for AI Legal Reasoning
Level 4: Domain Autonomous for AI Legal Reasoning
Level 5: Fully Autonomous for AI Legal Reasoning
Level 6: Superhuman Autonomous for AI Legal Reasoning

See **Figure A-1** for an overview chart showcasing the autonomous levels of AI Legal Reasoning as via columns denoting each of the respective levels.

See **Figure A-2** for an overview chart similar to Figure A-1 which alternatively is indicative of the autonomous levels of AI Legal Reasoning via the rows as depicting the respective levels (this is simply a reformatting of Figure A-1, doing so to aid in illuminating this variant perspective, but does not introduce any new facets or alterations from the contents as already shown in Figure A-1).

### 3.1.1 Level 0: No Automation for AI Legal Reasoning

Level 0 is considered the no automation level. Legal reasoning is carried out via manual methods and principally occurs via paper-based methods.

This level is allowed some leeway in that the use of say a simple handheld calculator or perhaps the use of a fax machine could be allowed or included within this Level 0, though strictly speaking it could be said that any form whatsoever of automation is to be excluded from this level.

### 3.1.2 Level 1: Simple Assistance Automation for AI Legal Reasoning

Level 1 consists of simple assistance automation for AI legal reasoning.

Examples of this category encompassing simple automation would include the use of everyday computer-based word processing, the use of everyday computer-based spreadsheets, the access to online legal documents that are stored and retrieved electronically, and so on.

By-and-large, today's use of computers for legal activities is predominantly within Level 1. It is assumed and expected that over time, the pervasiveness of automation will continue to deepen and widen, and eventually lead to legal activities being supported and within Level 2, rather than Level 1.

### 3.1.3 Level 2: Advanced Assistance Automation for AI Legal Reasoning

Level 2 consists of advanced assistance automation for AI legal reasoning.

Examples of this notion encompassing advanced automation would include the use of query-style Natural Language Processing (NLP), Machine Learning (ML) for case predictions, and so on.

Gradually, over time, it is expected that computer-based systems for legal activities will increasingly make use of advanced automation. Law industry technology that was once at a Level 1 will likely be refined, upgraded, or expanded to include advanced capabilities, and thus be reclassified into Level 2.

### 3.1.4 Level 3: Semi-Autonomous Automation for AI Legal Reasoning

Level 3 consists of semi-autonomous automation for AI legal reasoning.

Examples of this notion encompassing semi-autonomous automation would include the use of Knowledge-Based Systems (KBS) for legal reasoning, the use of Machine Learning and Deep Learning (ML/DL) for legal reasoning, and so on.

Today, such automation tends to exist in research efforts or prototypes and pilot systems, along with some commercial legal technology that has been infusing these capabilities too.

### 3.1.5 Level 4: Domain Autonomous for AI Legal Reasoning

Level 4 consists of domain autonomous computer-based systems for AI legal reasoning.

This level reuses the conceptual notion of Operational Design Domains (ODDs) as utilized in the autonomous vehicles and self-driving cars levels of autonomy, though in this use case it is being applied to the legal domain [17] [18] [20].

Essentially, this entails any AI legal reasoning capacities that can operate autonomously, entirely so, but that is only able to do so in some limited or constrained legal domain.

### 3.1.6 Level 5: Fully Autonomous for AI Legal Reasoning

Level 5 consists of fully autonomous computer-based systems for AI legal reasoning.

In a sense, Level 5 is the superset of Level 4 in terms of encompassing all possible domains as per however so defined ultimately for Level 4. The only constraint, as it were, consists of the facet that the Level 4 and Level 5 are concerning human intelligence and the capacities thereof. This is an important emphasis due to attempting to distinguish Level 5 from Level 6 (as will be discussed in the next subsection)

It is conceivable that someday there might be a fully autonomous AI legal reasoning capability, one that encompasses all of the law in all foreseeable ways, though this is quite a tall order and remains quite aspirational without a clear cut path of how this might one day be achieved. Nonetheless, it seems to be within the extended realm of possibilities, which is worthwhile to mention in relative terms to Level 6.

### 3.1.7 Level 6: Superhuman Autonomous for AI Legal Reasoning

Level 6 consists of superhuman autonomous computer-based systems for AI legal reasoning.

In a sense, Level 6 is the entirety of Level 5 and adds something beyond that in a manner that is currently ill-defined and perhaps (some would argue) as yet unknowable. The notion is that AI might ultimately exceed human intelligence, rising to become superhuman, and if so, we do not yet have any viable indication of what that superhuman intelligence consists of and nor what kind of thinking it would somehow be able to undertake.

Whether a Level 6 is ever attainable is reliant upon whether superhuman AI is ever attainable, and thus, at this time, this stands as a placeholder for that which might never occur. In any case, having such a placeholder provides a semblance of completeness, doing so without necessarily legitimatizing that superhuman AI is going to be achieved or not. No such claim or dispute is undertaken within this framework.

### 4.0 Turing Test Grid Integrating Autonomous Levels of AILR

### 4.1 Grid Indication of Levels of Autonomy (LoA) by Key Factors

In this section, the Turing Test key elements depicted in Section 2 are aligned into a grid that also contains the autonomous levels of AI Legal Reasoning which were described in Section 3.

**Figure B-1** provides an overview chart depicting the rows as the respective LoA AILR levels and the columns denoting the Turing Test elements. A row-by-row explanatory narrative is provided in the subsections below.

**Figure B-2** provides a similar overview chart of Figure B-1 but does so with the rows indicating the Turing Test key elements and the columns showcasing the AILR autonomous levels. This is simply an alternative perspective of Figure B-1 and does not introduce any new content or alterations from the contents depicted in Figure B-1. A row-by-row explanatory narrative is provided in the subsections below.

### 4.1.1 Level 0: No Automation for AI Legal Reasoning

As indicated in charts B-1 and B-2, Level 0 of the LoA AILR have an "n/a" (meaning not applicable) for each of the Turing Test key elements.

This designating of "n/a" is logically suitable for Level 0 since there is no autonomy associated with AILR at Level 0, therefore no relevancy in seeking to apply the Turing Test. Axiomatically, the Turing Test is inapplicable at Level 0. Any attempt to perform a Turing Test at Level 0 is inappropriate and unsuitable.

**Level 0**
- The Inquirer: **n/a**
- Human Participant: **n/a**
- AI-Based Legal Reasoner: **n/a**
- Queries of the Turing Test: **n/a**
- Answers to the Turing Test: **n/a**
- Rules of the Turing Test: **n/a**
- Potential Observers: **n/a**
- Conclusion Reached: **n/a**
- Reverse Turing Test: **n/a**

### 4.1.2 Level 1: Simple Assistance Automation for AI Legal Reasoning

As indicated in charts B-1 and B-2, Level 1 of the LoA AILR has an "n/a" (meaning not applicable) for each of the Turing Test key elements.

This designating of "n/a" is logically suitable for Level 1 since there is no autonomy associated with AILR at Level 1, therefore no relevancy in seeking to apply the Turing Test. Axiomatically, the Turing Test is inapplicable at Level 1. Any attempt to perform a Turing Test at Level 1 is inappropriate and unsuitable.

**Level 1**
- The Inquirer: **n/a**
- Human Participant: **n/a**
- AI-Based Legal Reasoner: **n/a**
- Queries of the Turing Test: **n/a**
- Answers to the Turing Test: **n/a**
- Rules of the Turing Test: **n/a**
- Potential Observers: **n/a**
- Conclusion Reached: **n/a**
- Reverse Turing Test: **n/a**

### 4.1.3 Level 2: Advanced Assistance Automation for AI Legal Reasoning

As indicated in charts B-1 and B-2, Level 2 of the LoA AILR has an "n/a" (meaning not applicable) for each of the Turing Test key elements.

This designating of "n/a" is logically suitable for Level 0 since there is no autonomy associated with AILR at Level 2, therefore no relevance in seeking to apply the Turing Test. Axiomatically, the Turing Test is inapplicable at Level 2. Any attempt to perform a Turing Test at Level 2 is inappropriate and unsuitable.

**Level 2**
- The Inquirer: **n/a**
- Human Participant: **n/a**
- AI-Based Legal Reasoner: **n/a**
- Queries of the Turing Test: **n/a**
- Answers to the Turing Test: **n/a**
- Rules of the Turing Test: **n/a**
- Potential Observers: **n/a**
- Conclusion Reached: **n/a**
- Reverse Turing Test: **n/a**

### 4.1.4 Level 3: Semi-Autonomous Automation for AI Legal Reasoning

As indicated in charts B-1 and B-2, Level 3 of the LoA AILR indicate several specific designations associated with the respective Turing Test elements.

Keep in mind that Level 3 is considered semi-autonomous, therefore situated partially in conventional automation and partially into autonomous capabilities. Since Level 3 is not defined as unqualified autonomy, there is no expectation that Level 3 AILR would be able to pass or succeed at the Turing Test. Nonetheless, it might be useful to administer the Turing Test as a means of gauging the extent of autonomous capabilities, along with being able to guide on what further advances might be needed to achieve Level 4 or higher.

For the inquirer, the preference is that an expert in legal reasoning would be utilized, rightfully so since the inquirer needs to be able to ask intelligent questions about the law, must be able to understand and assess the answers provided by the subjects participating, and must ultimately reach a conclusion about which is the human participant and which is the AI. The human participant should be an expert in the matters of legal reasoning being tested. The AI-based Legal Reasoner can consist of a minimal amount of AI legal reasoning capacity, having achieved a sufficient capacity to merit being categorized at Level 3. The queries of the Turing Test can be minimal in terms of the depth of exploration of legal reasoning, and likewise, the answers can be similarly of a minimal nature. Since this is viewed as a looser variant of the Turing Test, the rules of the matter can be minimal. Observers could be of an open nature and the conclusion reached by the inquirer is expected to be no greater than rated as "Notable" if the AILR can respond in a manner such that the Turing Test is considered as a pass. A Reverse Turing Test might be useful as a means to explore how to best further the AILR toward higher achievement in Level 3 or toward attainment of Level 4 or higher.

**Level 3**
- The Inquirer: **Expert Preferred**
- Human Participant: **Expert**
- AI-Based Legal Reasoner: **Minimal**
- Queries of the Turing Test: **Minimal**
- Answers to the Turing Test: **Minimal**
- Rules of the Turing Test: **Minimal**
- Potential Observers: **Open**
- Conclusion Reached: **Limited As "Notable"**
- Reverse Turing Test: **Useful But Not Substantive**

### 4.1.5 Level 4: Domain Autonomous for AI Legal Reasoning

As indicated in charts B-1 and B-2, Level 4 of the LoA AILR indicate several specific designations associated with the respective Turing Test elements.

Level 4 is considered autonomous with respect to a given legal domain. Therefore, this is considered an opportunity for a full-scale Turing Test in AILR, though restricted to the domain so specified. There is no expectation that the AILR would be able to pass or succeed outside the domain stipulated.

For the inquirer, an expert in the identified domain of legal reasoning would be utilized, rightfully so since the inquirer needs to be able to ask intelligent questions about the law in that domain, must be able to understand and assess the answers provided by the subjects participating as it relates to the domain, and must ultimately reach a conclusion about which is the human participant and which is the AI. The human participant should be an expert in the legal domain of legal reasoning being tested. The AI-based Legal Reasoner is to consist of a domain-specific AI legal reasoning capacity that fits the domain entailed, having achieved a sufficient capacity to merit being potentially categorized at Level 4.

The queries of the Turing Test should be bounded to the specific domain of legal reasoning, and likewise, the answers can be similarly bounded to the chosen domain. Since this is viewed as a full use of the Turing Test, the rules of the matter should be rigorously devised and applied. Observers would most likely be law specialists in the chosen domain and the conclusion reached by the inquirer is expected to be a domain-only pass if the AILR can respond in a manner such that the Turing Test is considered as succeeded. A Reverse Turing Test might be useful as a means to explore how to best further the AILR toward higher achievement in Level 4 or toward attainment of Level 5 or higher.

**Level 4**
- The Inquirer: **Expert in Domain**
- Human Participant: **Expert in Domain**
- AI-Based Legal Reasoner: **Domain Specific**
- Queries of the Turing Test: **Domain Specific**
- Answers to the Turing Test: **Domain Specific**
- Rules of the Turing Test: **Rigorous**
- Potential Observers: **Law Specialists**
- Conclusion Reached: **Domain-Only Pass in AILR**
- Reverse Turing Test: **Useful in Domain**

.

## 4.1.6 Level 5: Fully Autonomous for AI Legal Reasoning

As indicated in charts B-1 and B-2, Level 5 of the LoA AILR indicate several specific designations associated with the respective Turing Test elements.

Level 5 is considered autonomous with respect to all legal domains. Therefore, this is considered an opportunity for a full-scale Turing Test in AILR, being undertaken without any restrictions regarding the legal domains involved. The Turing Test should purposely seek to explore all legal domains since otherwise there would remain untested areas and any conclusion would be considered problematic.

For the inquirer, the likelihood is that one or more experts in the law would be utilized, rightfully so since an individual inquirer would seem unlikely to be able to encompass all legal domains, and the inquirer(s) need to be able to ask intelligent questions about the law in all legal domains, must be able to understand and assess the answers provided by the subjects participating as it relates to all legal domains, and must ultimately reach a conclusion about which is the human participant and which is the AI. The human participant might also consist of one or more experts due to the need to be able to respond across all legal domains and it seems unlikely that one individual could otherwise do so. The AI-based Legal Reasoner is to consist of an AI legal reasoning capacity that can be responsive across all legal domains, having achieved a sufficient capacity to merit being potentially categorized at Level 5.

The queries of the Turing Test should be bounded to the realm of law and require legal reasoning, and likewise, the answers are similarly bounded. Since this is viewed as a full use of the Turing Test, the rules of the matter should be rigorously devised and applied. Observers would most likely be law professionals across a variety of legal domains and the conclusion reached by the inquirer is expected to be a full pass if the AILR can respond in a manner such that the Turing Test is considered as succeeded. A Reverse Turing Test would likely be useful as a means to explore how to best further the AILR toward higher achievement in Level 5 or toward attainment of Level 6.

**Level 5**
- The Inquirer: **Multiple Experts**
- Human Participant: **Multiple Experts**
- AI-Based Legal Reasoner: **All Domains**
- Queries of the Turing Test: **All Domains**
- Answers to the Turing Test: **All Domains**
- Rules of the Turing Test: **Rigorous**
- Potential Observers: **Law Professionals**
- Conclusion Reached: **Full Pass in AILR**
- Reverse Turing Test: **Useful Overall**

### 4.1.7 Level 6: Superhuman Autonomous for AI Legal Reasoning

As indicated in charts B-1 and B-2, Level 6 of the LoA AILR indicate several specific designations associated with the respective Turing Test elements.

Level 6 is considered autonomous with respect to all legal domains. Therefore, this is considered an opportunity for a full-scale Turing Test in AILR, being undertaken without any restrictions regarding the legal domains involved. The Turing Test should purposely seek to explore all legal domains since otherwise there would remain untested areas and any conclusion would be considered problematic.

Level 6 poses a fundamental difficulty since it is as yet unknown as to what a superhuman capacity in the law might consist of, thus attempting to assess this capability via a Turing Test would seem challenging. Potentially, seemingly intractable legal questions that have stymied human legal reasoning might be utilized. Overall, it is unclear how those devising a test of an AI that is presumably at a heightened level of intellect could be suitably established since those creating the test are operating at a lower level of intellectual capacity. In any case, the Turing Test still might be viably applied and the nature of doing so is worthy of additional research, as mentioned in Section 5 of this paper.

For the inquirer, the likelihood is that one or more of the world's topmost experts in the law would be utilized, rightfully so since an individual inquirer would seem unlikely to be able to encompass all legal domains and since the attempt involves trying to challenge a superhuman AI capacity, and the inquirer(s) need to be able to ask hyper-intelligent questions about the law in all legal domains, must be able to understand and assess the answers provided by the subjects participating as it relates to all legal domains, and must ultimately reach a conclusion about which is the human participant and which is the AI. The human participant might also consist of one or more of the world's topmost experts due to the need to be able to respond across all legal domains and it seems unlikely that one individual could otherwise do so.

The AI-based Legal Reasoner is to consist of an AI legal reasoning capacity that can be responsive across all legal domains, having achieved a sufficient capacity to merit being potentially categorized at Level 6 and considered as presumably superhuman in capability. The queries of the Turing Test should be bounded to the realm of law and require legal reasoning, and likewise, the answers are similarly bounded. Since this is viewed as a full use of the Turing Test, the rules of the matter should be rigorously devised and applied. Observers would most likely be both those versed in the law and those non-law observers interested in the superhuman capacity overall, and the conclusion reached by the inquirer is expected to be an exemplary pass if the AILR can respond in a manner such that the Turing Test is considered as succeeded. A Reverse Turing Test would likely be useful as a means to explore how to best further the AILR toward higher achievement in Level 6.

> **Level 6**
> - The Inquirer: **Topmost Experts**
> - Human Participant: **Topmost Experts**
> - AI-Based Legal Reasoner: **Domain Plus**
> - Queries of the Turing Test: **Domain Plus**
> - Answers to the Turing Test: **Domain Plus**
> - Rules of the Turing Test: **Rigorous**
> - Potential Observers: **Law & Non-Law**
> - Conclusion Reached: **Exemplary Pass in AILR**
> - Reverse Turing Test: **Useful Overall**

### 4.2 Grid Indication of Turing Test Key Elements by Levels of Autonomy (LoA)

The next subsections showcase the Turing Test key factors as at-a-glance for each factor, listing the designations that have been postulated for each of the LoA AILR levels.

Narrative discussion about these facets has already been covered in the prior Subsection 4.1 and thus it is not necessary to repeat it in this subsection (refer to the prior subsections as needed).

#### 4.2.1 Turing Test "The Inquirer" by LoA

For a narrative discussion about the "The Inquirer" for each of the LoA AILR levels, see the preceding subsections. This list shown here provides a convenience of indication and is also portrayed on charts B-1 and B-2.

**The Inquirer**
- Level 0: **n/a**
- Level 1: **n/a**
- Level 2: **n/a**
- Level 3: **Expert Preferred**
- Level 4: **Expert in Domain**
- Level 5: **Multiple Experts**
- Level 6: **Topmost Experts**

### 4.2.2 Turing Test "Human Participant" by LoA

For a narrative discussion about the "Human Participant" for each of the LoA AILR levels, see the preceding subsections. This list shown here provides a convenience of indication and is also portrayed on charts B-1 and B-2.

**Human Participant**
- Level 0: **n/a**
- Level 1: **n/a**
- Level 2: **n/a**
- Level 3: **Expert**
- Level 4: **Expert in Domain**
- Level 5: **Multiple Experts**
- Level 6: **Topmost Experts**

### 4.2.3 Turing Test "AI-Based Legal Reasoner" LoA

For a narrative discussion about the "AI-Based Legal Reasoner" for each of the LoA AILR levels, see the preceding subsections. This list shown here provides a convenience of indication and is also portrayed on charts B-1 and B-2.

**AI-Based Legal Reasoner**
- Level 0: **n/a**
- Level 1: **n/a**
- Level 2: **n/a**
- Level 3: **Minimal**
- Level 4: **Domain Specific**
- Level 5: **All Domains**
- Level 6: **Domains Plus**

### 4.2.4 Turing Test "Queries of the Turing Test" by LoA

For a narrative discussion about the "Queries of the Turing Test" for each of the LoA AILR levels, see the preceding subsections. This list shown here provides a convenience of indication and is also portrayed on charts B-1 and B-2.

**Queries of the Turing Test**
- Level 0: **n/a**
- Level 1: **n/a**
- Level 2: **n/a**
- Level 3: **Minimal**
- Level 4: **Domain Specific**
- Level 5: **All Domains**
- Level 6: **Domain Plus**

### 4.2.5 Turing Test "Answers to the Turing Test" by LoA

For a narrative discussion about the "Answers to the Turing Test" for each of the LoA AILR levels, see the preceding subsections. This list shown here provides a convenience of indication and is also portrayed on charts B-1 and B-2.

**Answers to the Turing Test**
- Level 0: **n/a**
- Level 1: **n/a**
- Level 2: **n/a**
- Level 3: **Minimal**
- Level 4: **Domain Specific**
- Level 5: **All Domains**
- Level 6: **Domain Plus**

### 4.2.6 Turing Test "Rules of the Turing Test" by LoA

For a narrative discussion about the "Rules of the Turing Test" for each of the LoA AILR levels, see the preceding subsections. This list shown here provides the convenience of indication and is also portrayed on charts B-1 and B-2.

**Rules of the Turing Test**
- Level 0: **n/a**
- Level 1: **n/a**
- Level 2: **n/a**
- Level 3: **Minimal**
- Level 4: **Rigorous**
- Level 5: **Rigorous**
- Level 6: **Rigorous**

### 4.2.7 Turing Test "Potential Observers" LoA

For a narrative discussion about the "Potential Observers" for each of the LoA AILR levels, see the preceding subsections. This list shown here provides a convenience of indication and is also portrayed on charts B-1 and B-2.

**Potential Observers**
- Level 0: **n/a**
- Level 1: **n/a**
- Level 2: **n/a**
- Level 3: **Open**
- Level 4: **Law Specialists**
- Level 5: **Law Professionals**
- Level 6: **Law & Non-Law**

### 4.2.8 Turing Test "Conclusions Reached" by LoA

For a narrative discussion about the "Conclusions Reached" for each of the LoA AILR levels, see the preceding subsections. This list shown here provides a convenience of indication and is also portrayed on charts B-1 and B-2.

**Conclusion Reached**
- Level 0: **n/a**
- Level 1: **n/a**
- Level 2: **n/a**
- Level 3: **Limited As "Notable"**
- Level 4: **Domain-Only Pass in AILR**
- Level 5: **Full Pass in AILR**
- Level 6: **Exemplary Pass in AILR**

### 4.2.9 Turing Test "Reverse Turing Test" by LoA

For a narrative discussion about the "Reverse Turing Test" for each of the LoA AILR levels, see the preceding subsections. This list shown here provides a convenience of indication and is also portrayed on charts B-1 and B-2.

**Reverse Turing Test**
- Level 0: **n/a**
- Level 1: **n/a**
- Level 2: **n/a**
- Level 3: **Useful But Not Substantive**
- Level 4: **Useful in Domain**
- Level 5: **Useful Overall**
- Level 6: **Useful Overall**

## 5.0 Additional Considerations and Future Research

The grid depicted in Figure B-1 and Figure B-2 is a strawman variant, meaning that the indications shown are an initial populating of the grid. Additional research is needed to explore the designations and ascertain whether the initial indications might be advisedly changed or possibly transformed into some other kind of designations, such as numeric scores or weights.

Another aspect of additional research involves the Turing Test key elements that are utilized in this strawman variant. There are other ways to portray the elements, along with the possibility of adding elements or possibly opting to excise some of the elements from the grid. Research on such modifications is encouraged.

As a final point, there are potentially greater questions that arise from the grid, alluded to earlier in the discussion of the prior sections, entailing what actions would be taken if indeed AILR can achieve the autonomous levels of Level 4, Level 5, and Level 6. There remain many such open issues, each deserving of suitable attention.

This paper has proposed a variant of the Turing Test that is customized for specific use in the AILR realm, including depicting how this famous "gold standard" of AI fulfillment can be robustly applied across the autonomous levels of AI Legal Reasoning. Such an instrument can aid in addressing the open question underlying how we will know when AILR has achieved autonomous capacities.

**About the Author**

Dr. Lance Eliot is the Chief AI Scientist at Techbrium Inc. and a Stanford Fellow at Stanford University in the CodeX: Center for Legal Informatics. He previously was a professor at the University of Southern California (USC) where he headed a multi-disciplinary and pioneering AI research lab. Dr. Eliot is globally recognized for his expertise in AI and is the author of highly ranked AI books and columns.

**Figure A-1**

# AI & Law: Levels of Autonomy For AI Legal Reasoning (AILR)

| Level | Descriptor | Examples | Automation | Status |
|---|---|---|---|---|
| 0 | No Automation | Manual, paper-based (no automation) | None | De Facto - In Use |
| 1 | Simple Assistance Automation | Word Processing, XLS, online legal docs, etc. | Legal Assist | Widely In Use |
| 2 | Advanced Assistance Automation | Query-style NLP, ML for case prediction, etc. | Legal Assist | Some In Use |
| 3 | Semi-Autonomous Automation | KBS & ML/DL for legal reasoning & analysis, etc. | Legal Assist | Primarily Prototypes & Research Based |
| 4 | AILR Domain Autonomous | Versed only in a specific legal domain | Legal Advisor (law fluent) | None As Yet |
| 5 | AILR Fully Autonomous | Versatile within and across all legal domains | Legal Advisor (law fluent) | None As Yet |
| 6 | AILR Superhuman Autonomous | Exceeds human-based legal reasoning | Supra Legal Advisor | Indeterminate |

*Figure 1: AI & Law - Autonomous Levels by Rows*     *Source Author: Dr. Lance B. Eliot*

V1.3

**Figure A-2**

| | Level 0 | Level 1 | Level 2 | Level 3 | Level 4 | Level 5 | Level 6 |
|---|---|---|---|---|---|---|---|
| **Descriptor** | No Automation | Simple Assistance Automation | Advanced Assistance Automation | Semi-Autonomous Automation | AILR Domain Autonomous | AILR Fully Autonomous | AILR Superhuman Autonomous |
| **Examples** | Manual, paper-based (no automation) | Word Processing, XLS, online legal docs, etc. | Query-style NLP, ML for case prediction, etc. | KBS & ML/DL for legal reasoning & analysis, etc. | Versed only in a specific legal domain | Versatile within and across all legal domains | Exceeds human-based legal reasoning |
| **Automation** | None | Legal Assist | Legal Assist | Legal Assist | Legal Advisor (law fluent) | Legal Advisor (law fluent) | Supra Legal Advisor |
| **Status** | De Facto – In Use | Widely In Use | Some In Use | Primarily Prototypes & Research-based | None As Yet | None As Yet | Indeterminate |

*Figure 2: AI & Law - Autonomous Levels by Columns*   *Source Author: Dr. Lance B. Eliot*   V1.3

**Figure B-1**

| | Level 0 | Level 1 | Level 2 | Level 3 | Level 4 | Level 5 | Level 6 |
|---|---|---|---|---|---|---|---|
| | \multicolumn{7}{c}{Turing Test and Autonomous Levels of AI Legal Reasoning (AILR)} | | | | | | |
| Descriptor | No Automation | Simple Assistance Automation | Advanced Assistance Automation | Semi-Autonomous Automation | AILR Domain Autonomous | AILR Fully Autonomous | AILR Superhuman Autonomous |
| The Inquirer | n/a | n/a | n/a | Expert Preferred | Expert in Domain | Multiple Experts | Topmost Experts |
| Human Participant | n/a | n/a | n/a | Expert | Expert in Domain | Multiple Experts | Topmost Experts |
| AI-Based Legal Reasoner | n/a | n/a | n/a | Minimal | Domain Specific | All Domains | Domains Plus |
| Queries of the Turing Test | n/a | n/a | n/a | Minimal | Domain Specific | All Domains | Domains Plus |
| Answers to the Turing Test | n/a | n/a | n/a | Minimal | Domain Specific | All Domains | Domains Plus |
| Rules of the Turing Test | n/a | n/a | n/a | Minimal | Rigorous | Rigorous | Rigorous |
| Potential Observers | n/a | n/a | n/a | Open | Law Specialists | Law Professionals | Law & Non-Law |
| Conclusion Reached | n/a | n/a | n/a | Limited As "Notable" | Domain-Only Pass in AILR | Full Pass In AILR | Exemplary Pass in AILR |
| Reverse Turing Test | n/a | n/a | n/a | Useful But Not Substantive | Useful In Domain | Useful Overall | Useful Overall |

*Strawman Variant*

Figure 1: AI & Law – Turing Test and LoA AILR by Columns    Source Author: Dr. Lance B. Eliot    V1.3

**Figure B-2**

## Turing Test and Levels of Autonomy For AI Legal Reasoning (AILR)

| Level | Descriptor | The Inquirer | Human Participant | AI-Based Legal Reasoner | Queries of the Turing Test | Answers to the Turing Test | Rules of The Turing Test | Potential Observers | Conclusion Reached | Reverse Turing Test |
|---|---|---|---|---|---|---|---|---|---|---|
| 0 | No Automation | n/a | n/a | n/a | n/a | n/a | n/a | n/a | n/a | n/a |
| 1 | Simple Assistance Automation | n/a | n/a | n/a | n/a | n/a | n/a | n/a | n/a | n/a |
| 2 | Advanced Assistance Automation | n/a | n/a | n/a | n/a | n/a | n/a | n/a | n/a | n/a |
| 3 | Semi-Autonomous Automation | Expert Preferred | Expert | Minimal | Minimal | Minimal | Minimal | Open | Limited As "Notable" | Useful But Not Substantive |
| 4 | AILR Domain Autonomous | Expert in Domain | Expert in Domain | Domain Specific | Domain Specific | Domain Specific | Rigorous | Law Specialists | Domain-Only Pass in AILR | Useful in Domain |
| 5 | AILR Fully Autonomous | Multiple Experts | Multiple Experts | All Domains | All Domains | All Domains | Rigorous | Law Professionals | Full Pass in AILR | Useful Overall |
| 6 | AILR Superhuman Autonomous | Topmost Experts | Topmost Experts | Domain Plus | Domain Plus | Domain Plus | Rigorous | Law & Non-Law | Exemplary Pass in AILR | Useful Overall |

Figure 2: AI & Law – Turing Test and LoA AILR by Rows    *Strawman Variant*    Source Author: Dr. Lance B. Eliot    V1.3

**Figure B-3**

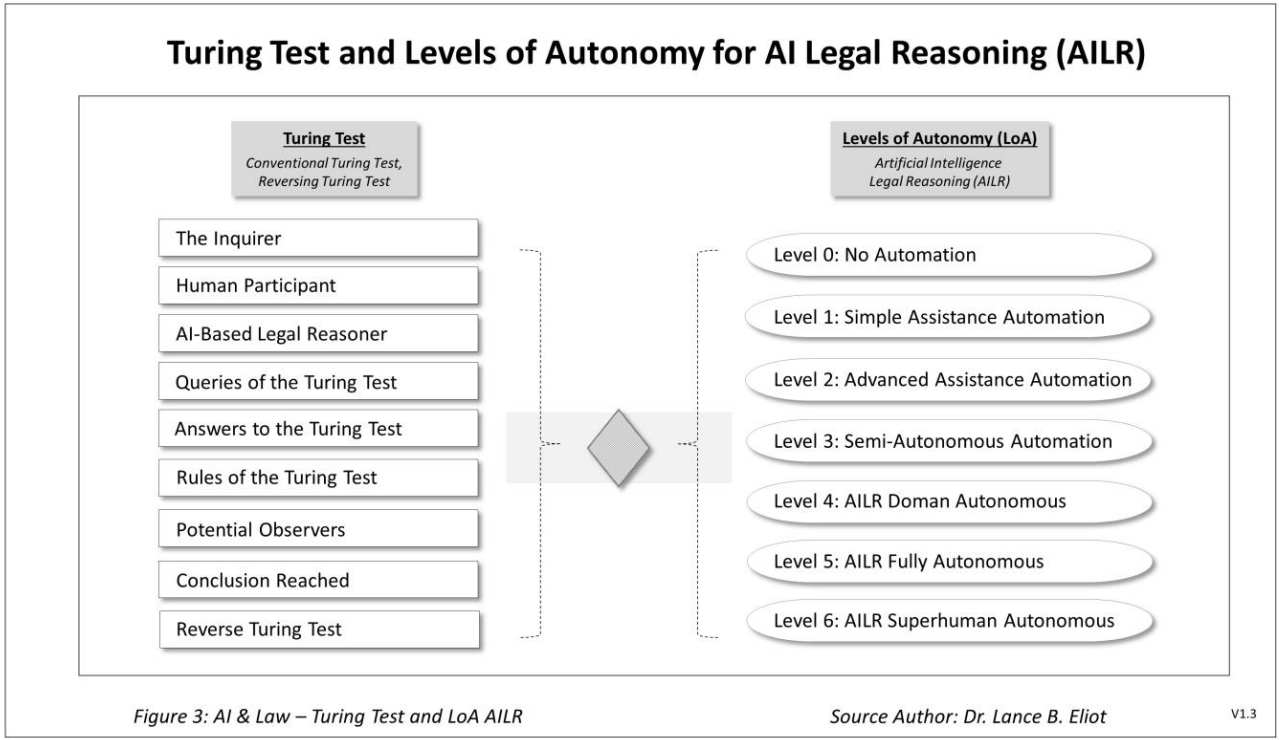

Figure 3: AI & Law – Turing Test and LoA AILR   Source Author: Dr. Lance B. Eliot   V1.3

**Figure B-4**

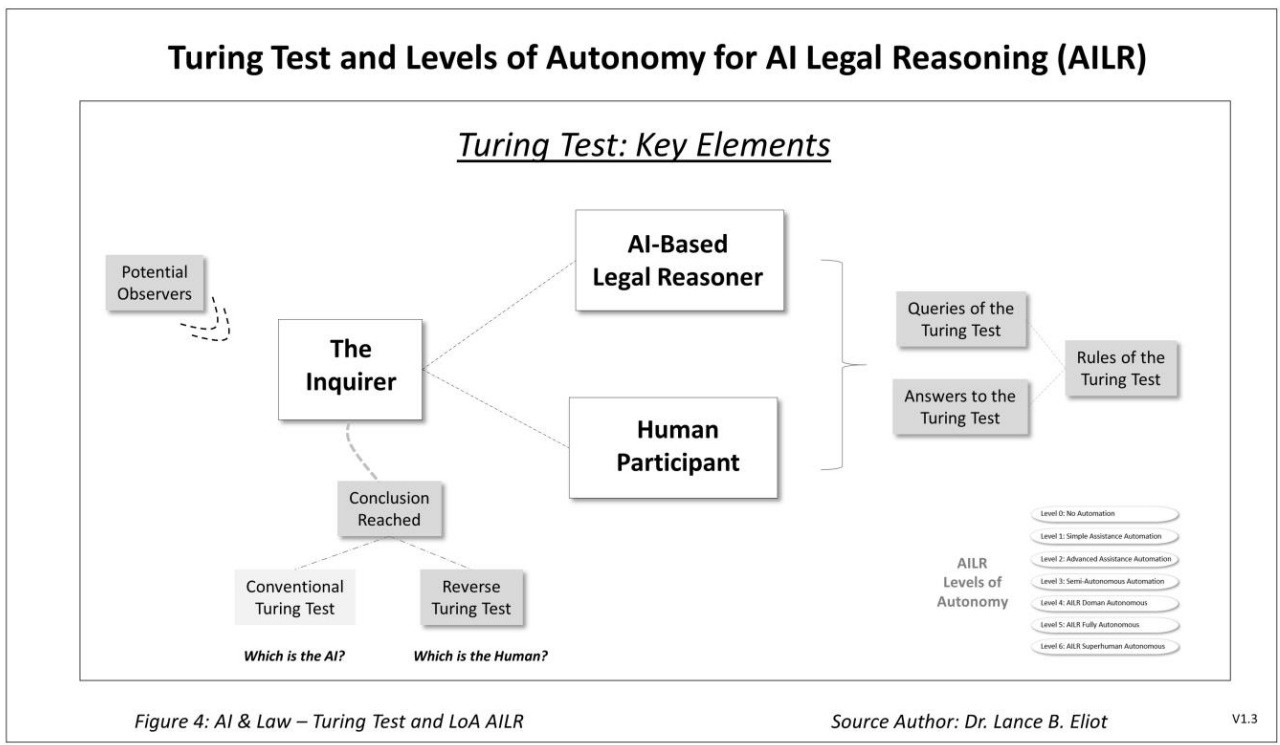